\documentclass[fleqn,usenatbib]{mnras}

\usepackage{newtxtext,newtxmath}

\usepackage[T1]{fontenc}
\usepackage{ae,aecompl}

\usepackage{graphicx}	
\usepackage{amsmath}	
\usepackage{amssymb}	

\usepackage[figure,figure*]{hypcap}

\title[UVIT view of ram-pressure stripping in action]{UVIT view of ram-pressure stripping in action: Star formation in the stripped gas of the GASP jellyfish galaxy JO201 in Abell 85}

\author[K. George et al.]{
K. George$^{1}$,\thanks{E-mail: koshy@iiap.res.in}
B. M. Poggianti$^2$,
M. Gullieuszik$^2$,
G. Fasano$^2$,
C. Bellhouse$^{3,4}$,J. Postma$^{5}$,
\newauthor  
A. Moretti$^2$, Y. Jaff\'e$^6$, B. Vulcani$^{2,7}$,D. Bettoni$^2$,
J. Fritz$^8$,   
P. C{\^o}t{\'e}$^{9}$,
S. K. Ghosh$^{10,11}$,
\newauthor  
J. B. Hutchings$^{9}$,
R. Mohan$^{1}$,
P. Sreekumar$^{1}$,
C. S. Stalin$^{1}$,
A. Subramaniam$^{1}$,
\newauthor
S.N. Tandon$^{1,12}$
\\
$^{1}$Indian Institute of Astrophysics, Koramangala II Block, Bangalore, India\\
$^{2}$INAF-Astronomical Observatory of Padova   
vicolo dell'Osservatorio 5   
35122 Padova, Italy\\
$^{3}$European Southern Observatory, Alonso de Cordova 3107,
  Vitacura, Casilla 19001, Santiago de Chile, Chile\\
$^{4}$University of Birmingham School of Physics and Astronomy,
  Edgbaston, Birmingham, England\\
$^{5}$University of Calgary, Calgary, Alberta, Canada\\
$^{6}$Instituto de F\'isicay Astronom\'ia, Universidad de Valpara\'iso, Gran Breta\~na 1111, Valpara\'iso, Chile\\  
$^{7}$School of Physics, The University of Melbourne, Swanston St \&
  Tin Alley Parkville, VIC 3010, Australia\\ 
$^{8}$Instituto de Radioastronomia y Astrofisica, UNAM, Campus
  Morelia, A.P. 3-72, C.P. 58089, Mexico\\ 
$^{9}$National Research Council of Canada, Herzberg Astronomy and Astrophysics Research Centre, Victoria, Canada\\
$^{10}$National Centre for Radio Astrophysics, Pune, India\\
$^{11}$Tata Institute of Fundamental Research, Mumbai, India\\
$^{12}$Inter-University Center for Astronomy and Astrophysics, Pune, India}
\date{Accepted XXX. Received YYY; in original form ZZZ}

\pubyear{2018}

\begin{document}
\label{firstpage}
\pagerange{\pageref{firstpage}--\pageref{lastpage}}
\maketitle

\begin{abstract}
Jellyfish are cluster galaxies that experience strong ram-pressure effects that strip their gas. Their H$\alpha$ images reveal ionized gas tails up to 100 kpc, which could be hosting ongoing star formation. Here we  report the ultraviolet (UV) imaging observation of the jellyfish galaxy JO201 obtained at a spatial resolution $\sim$ 1.3 kpc. The intense burst of star formation happening in the tentacles is the focus of the present study. JO201 is the "UV-brightest cluster galaxy" in Abell 85 ($z \sim$ 0.056) with knots and streams of star formation in the ultraviolet. We identify star forming knots both in the stripped gas and in the galaxy disk and compare the UV features with the ones traced by H$\alpha$ emission. Overall, the two emissions remarkably correlate, both in the main body and along the tentacles. Similarly, also the star formation rates of individual knots derived from the extinction-corrected FUV emission agree with those derived from the H$\alpha$ emission and range from $\sim$ 0.01 -to- 2.07 $M_{\odot} \, yr^{-1}$. The integrated star formation rate from FUV flux is $\sim$ 15 $M_{\odot} \, yr^{-1}$. The unprecedented deep UV imaging study of the jellyfish galaxy JO201 shows clear signs of extraplanar star-formation activity due to a recent/ongoing gas stripping event.

\end{abstract}

\begin{keywords}
galaxies: clusters: intracluster medium, galaxies: star formation
\end{keywords}



\section{Introduction} \label{sec:intro}

Galaxies in the local Universe follow a bimodal distribution in the optical color-magnitude diagram with a red sequence populated by old, red galaxies and a blue cloud with young actively star-forming blue galaxies \citep{Visvanathan_1977,Baldry_2004}. Galaxies on the red sequence are mostly of early-type (E/S0) morphology and host little cold gas and dust, whereas galaxies on the blue cloud have late-type (spirals) morphology and usually host an abundant cold gas content. A significant fraction of red sequence galaxies are observed to be located in the densest regions of the local Universe like  the cores of massive galaxy clusters. The blue cloud galaxies are found mostly in the low density regions of the Universe, i.e.\ in the field environment, as well as in cluster outskirts.\\

The change in the morphology and the star formation properties of galaxies with changing galaxy density is now believed to be influenced by the rapid decline in star formation as galaxies fall into the dense cluster environment from the field. The morphology-density relation, the Butcher-Oemler effect and the high occurrence of blue star-forming galaxies in the cluster outskirts are the observational support for this galaxy transformation in dense environments. \citep{Dressler_1980,Butcher_1984,Dressler_2009,Mahajan_2012,Fasano_2015}. There are multiple processes, such as strangulation, harassment and ram pressure stripping, which can act alone or in combination to convert a star-forming galaxy into a non star-forming one when they experience the cluster environment. \\

Ram pressure is the main mechanism in quenching star formation in cluster galaxies. The intra-cluster medium is composed of hot X-ray emitting plasma with temperatures in the range $10^7-10^8$ K and electron density in the range $10^{-4} - 10^{-2} cm^{-3}$ contained in a virial radius \citep{Sarazin_1986,Fabian_1994}. When a galaxy falls into the intra-cluster medium, its interstellar medium experiences a force in the opposite direction of the relative motion. The cold gas gets stripped from the disk of the fast moving spiral galaxies through the process of ram pressure stripping \citep{Gunn_1972}. The in-falling gas rich spiral galaxies thus go through a  phase of morphological transformation, partly or even fully due to gas removal processes. The stripping of gas quenches the star formation and transforms the galaxies into passively evolving red sequence systems \citep{Dressler_1997,Poggianti_1999,Dressler_2013}.\\ 

Gas-rich spiral galaxies have spatially extended loose gas halos and tightly bound disk gas, both of which are subjected to the harsh impact of cluster in-fall \citep{Bekki_2009}. The observations and simulations of in-falling galaxies have given ample evidence for the presence of stripped neutral and molecular hydrogen in the opposite direction to the orbital velocity vector. \citep{Haynes_1984,Cayatte_1990,Kenney_2004,Chung_2009,Vollmer_2001,Yamagami_2011,Serra_2013,Jaffe_2015,Yoon_2017,Tonnesen_2009,Tonnesen_2012,Jachym_2014,Verdugo_2015,Jachym_2017,Moretti_2018}. The gas gets stripped but also shock compressed, and this can trigger pockets of intense star formation. Galaxies undergoing strong ram-pressure events can sometimes be identified in optical observations due to the existence of tentacles of debris material resembling a jellyfish.  The optical and  $\mathrm{H}{\alpha}$ observations of galaxies falling into nearby galaxy clusters have shown disturbed  $\mathrm{H}{\alpha}$ emission with undisturbed stellar disks  \citep{Kenney_1999,Yoshida_2008,Hester_2010,Kenney_2014}. The triggered star formation in the stripped gas appear as tentacles and give the galaxy a visual appearance of jellyfish morphology \citep{Owen_2006,Cortese_2007,Owers_2012,Fumagalli_2014,Ebeling_2014,Rawle_2014,Poggianti_2016,Bellhouse_2017,Gullieuszik_2017,Boselli_2018}. Star formation happening in the compressed gas within the galaxy and in the ejected gas makes these galaxies bright in the ultraviolet. Young, massive stars (O,B,A spectral types) emit the bulk of radiation in the ultraviolet (UV) region of the spectral energy distribution and hence UV can be used as a direct probe to study ongoing star formation. The nature of the intense star formation in the main body of the galaxy and the debris material of the jellyfish galaxies can thus be directly studied using ultraviolet observations. \\

We present here the ultraviolet study of the jellyfish galaxy JO201 undergoing extreme ram-pressure stripping in the massive galaxy cluster Abell 85. {\sl The aim of the present study is to identify the sites of intense star formation in the stripped gas and in the galaxy disk and estimate the star formation rates.} JO201 is taken from the sample of 419 (344 cluster and 75 field) jellyfish candidates of \citet{Poggianti_2016} and is one of the most striking cases of ram-pressure stripping in action (see \citet{Bellhouse_2017, Poggianti_2017,Jaffe_2018}. The galaxy has a spiral morphology with tails of material to one side in the optical images and total stellar mass $\sim 6 \times 10^{10} M_{\odot}$ for a Salpeter IMF between 0.1 and 100 $M_{\odot}$ \citep{Bellhouse_2017,Salpeter_1955}. (The total stellar mass is computed for a Chabrier IMF in \citet{Bellhouse_2017}. Here we compute the value for a Salpeter IMF). The galaxy is falling into the cluster from the back with a slight inclination from the line-of-sight directed to the west: this explains the jellyfish morphology with projected tails pointing towards the east, in the direction of the brightest cluster galaxy (BCG). The line-of-sight velocity of JO201 (3363.7 km/s) is very high with respect to the mean velocity of the Abell 85 galaxy cluster. In what follows we compare the UV and $\mathrm{H}{\alpha}$ emission across JO201. We discuss the observations in section 2, and present the results in section 3. We summarize the key findings from the study in section 4. Throughout this paper we adopt a Salpeter 0.1-100 $M_{\odot}$ initial mass function, and a concordance $\Lambda$ CDM cosmology with $H_{0} = 70\,\mathrm{km\,s^{-1}\,Mpc^{-1}}$, $\Omega_{\rm{M}} = 0.3$, $\Omega_{\Lambda} = 0.7$.

\begin{table}
\caption{\label{t7} Log of UVIT observations of Abell 85 galaxy cluster.}
\centering
\label{galaxy details}
\tabcolsep=0.05cm
\begin{tabular}{ccccc} 
\hline
\hline
Channel & Filter & $\lambda_{mean}$({\AA})  &  $\delta$$\lambda$({\AA})  & Int:time(s)  \\
\hline
FUV  & F148W       &  1481  &   500     &  15429\\
NUV  & N242W     &  2418  &   785     &  18326 \\
\hline
\end{tabular}

\end{table}

\section{Observations, Data \& Analysis} \label{sec:style}
 The galaxy JO201 was observed at optical wavelengths as part of the WINGS and OmegaWINGS surveys \citep{Fasano_2006,Gullieuszik_2015,Moretti_2017} and with MUSE on the VLT under the programme GASP (GAs Stripping Phenomena in galaxies with MUSE) aimed at investigating the gas removal process in galaxies using the spatially resolved integral field unit spectrograph MUSE \citep{Poggianti_2017,Bellhouse_2017}. JO201 (RA: 00:41:30.325, Dec: - 09:15:45.96) belongs to the massive galaxy cluster Abell 85 ($M_{200}=1.58 \times 10^{15} \mathrm{M}_{\odot}$) at a redshift $\sim 0.056$ \citep{Moretti_2017}. The corresponding luminosity distance is $\sim$ 250  Mpc and the angular scale of 1" on the sky corresponds to 1.087 kpc at the galaxy cluster rest frame. \\

The jellyfish galaxy JO201 was observed with the ultra-violet imaging telescope (UVIT) onboard the Indian multi wavelength astronomy satellite ASTROSAT \citep{Agrawal_2006}. The UVIT consists of twin telescopes, a FUV (130-180nm) telescope and a NUV (200-300nm),VIS (320-550nm) telescope which operates with a dichroic beam splitter. The telescopes are of 38cm diameter and generate circular images over a 28$'$ diameter field simultaneously in all three channels \citep{Kumar_2012}. 
There are options for a set of narrow and broad band filters, out of which we used the NUV N242W and FUV F148W  filters for NUV and FUV imaging observations. Table 1 gives details on the UVIT observations of the Abell 85 galaxy cluster. We note that there are GALEX observations of JO201 with an integration time of 25ks in NUV and 2.5ks in FUV channel (also see \citet{Venkatapathy_2017}). The UVIT observations for JO201 are at an angular resolution of $\sim$ 1\farcs2  for the NUV  and  $\sim$ 1\farcs4  for the FUV channels, while the GALEX resolution is $\sim$ 4-5 arcsec \footnote{UVIT NUV N242W and FUV F148W filters have similar bandpass to GALEX NUV and FUV filters.} \citep{Annapurni_2016,Tandon_2016}. The NUV and FUV images are corrected for distortion \citep{Girish_2017}, flat field and satellite drift using the software CCDLAB  \citep{Postma_2017}. The images from 10 orbits are coadded to create the master image. The astrometric calibration is performed using the {\tt astrometry.net} package  where solutions are performed using USNO-B catalog \citep{Lang_2010}. The photometric calibration is done using the zero point values generated for photometric calibration stars as described in  \citet{Tandon_2017}. Note that magnitudes are in AB system. \\

\begin{figure*}\centering
\includegraphics[width=1\textwidth]{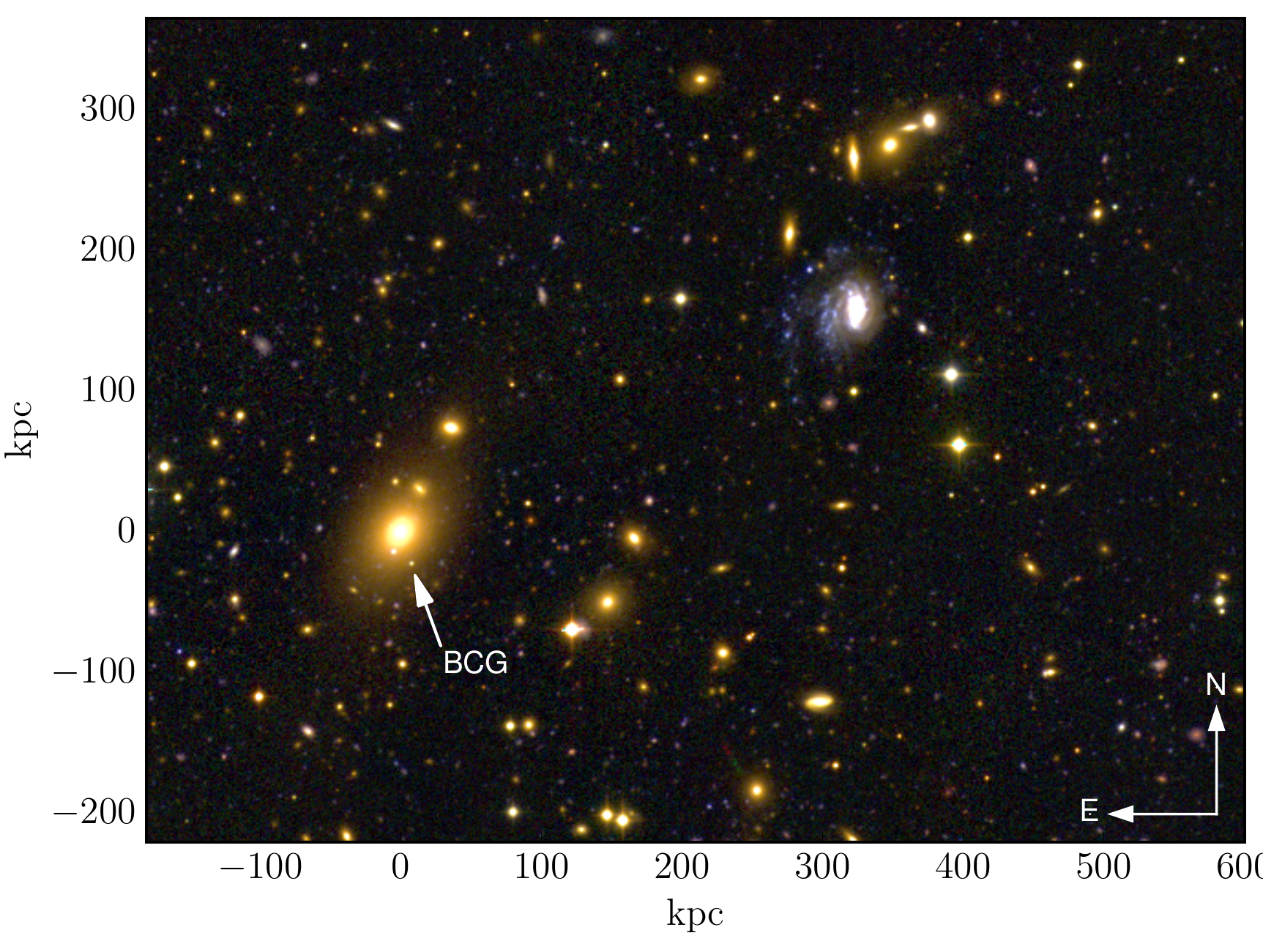}
\caption{Color-composite image of the Abell 85 galaxy cluster field.  The image is made from combining NUV (colored blue) and optical $B,V$ filter band pass images. The jellyfish galaxy is prominent in NUV as evident from the enhanced blue color from JO201. The position of the brightest cluster galaxy (BCG) is shown.The image is of size $\sim$ 12.0$'$ $\times$ 9.0$'$.}\label{figure:A85comp}
\end{figure*}

Abell 85 galaxy cluster imaging observations in the $B$ and $V$ bands were taken as part of the WINGS survey \citep{Varela_2009} and the galaxy JO201 was observed with the MUSE integral-field spectrograph mounted on the ESO Very large Telescope as part of the GASP program, with photometric conditions and image quality of $\sim$ 0\farcs7 FWHM, as described in detail in \citet{Bellhouse_2017}. MUSE has a 1$\arcmin$ $\times$ 1$\arcmin$ field of view and JO201 was covered with two MUSE pointings. The MUSE observations cover the stripped tails of the galaxy and confirm that JO201 is indeed a jellyfish galaxy undergoing intense ram-pressure stripping by the intra-cluster medium, close to the core of the massive cluster Abell 85. These data reveal extended $\mathrm{H}{\alpha}$ emission out to $\sim$ 60kpc from the stellar disk of the galaxy with kinematics indicative of significant stripping in the line-of-sight direction \citep{Bellhouse_2017}.The $\mathrm{H}{\alpha}$ emission line flux map of JO201 from MUSE (which we call $\mathrm{H}{\alpha}$ image as described in \citet{Bellhouse_2017}) is used in this study for comparison with UV imaging data. We extracted a region around JO201 from the NUV and FUV images and recentered the astrometric solution to match point sources from WINGS catalog. The UV and $\mathrm{H}{\alpha}$ images are then assigned to the same astrometric reference frame with an accuracy better than 1$\arcsec$. A color composite image of the Abell 85 cluster central region including both JO201 and the Brightest Cluster Galaxy created using NUV, $B$ and $V$ band images is shown in Figure \ref{figure:A85comp}. The red and green colors correspond to the flux from the $B$ and $V$ images from the WINGS survey and the NUV image from UVIT is shown in blue color.

\section{Results}\label{sec:Results}

\subsection{Ultraviolet imaging}

\begin{figure}
\centerline{\includegraphics[width=0.53\textwidth]{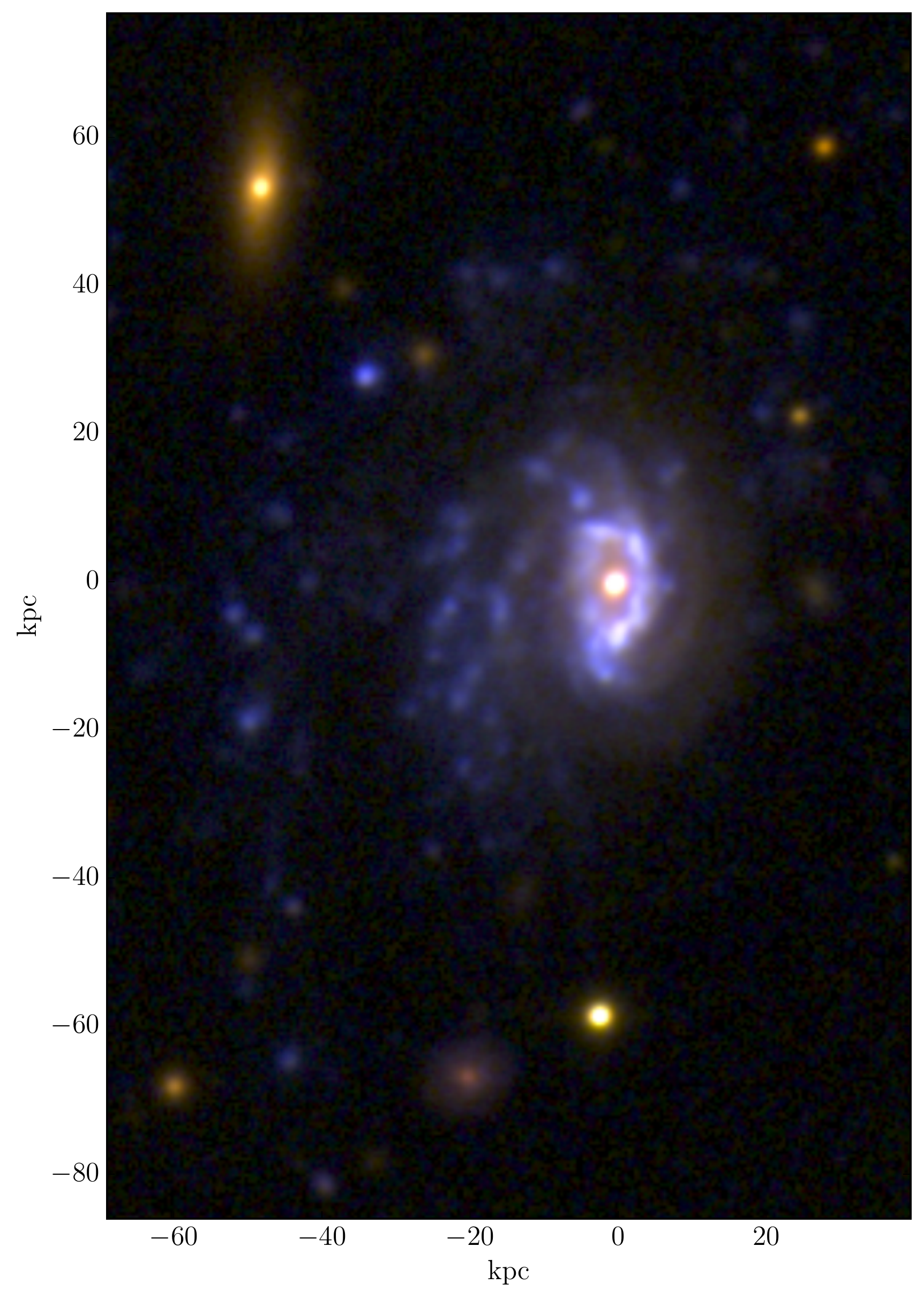}}
\caption{The NUV and optical composite image of J0201. The NUV image is from observations with UVIT and optical imaging from WINGS survey. The image is made from combining NUV (colored blue) and optical $B,V$ filter band pass images (colored red and green). Note the diffuse emission and the knots of star formation in the stripped material from the galaxy. The bright source at the center is powered by an active galactic nucleus. The image is of size $\sim$ 100$"$ $\times$ 150$"$.}\label{figure:JO201comp}
\end{figure}

The UVIT 28$'$ field of view corresponds to $\sim$ 1.83 Mpc at the Abell 85 galaxy cluster rest frame. The NUV and the FUV images display similar morphological features, but the NUV image has higher angular resolution than the FUV image and therefore we use the former to identify star forming regions in JO201 and quantify the flux within the detected knots. The NUV image of a region centered on JO201 shown in Figure \ref{figure:JO201comp} displays a wealth of information, which include low surface brightness features outside of the galaxy disk, knots on the galaxy disk and knots outside of the galaxy in the intergalactic space that seem to follow the spiral pattern from the galaxy disk (see Fig. 8 in \citet{Bellhouse_2017}), possibly suggesting unwinding of spiral arms.  The NUV emission from the main body and the tentacles of JO201 appears to be clumpy in nature with knots having flux due to emission from star formation. The bright source at the center of JO201 corresponds to the AGN identified by \citet{Poggianti_2017} (see also Bellhouse et al. in prep).

\subsection{Combining Ultraviolet and $\mathrm{H}{\alpha}$ }
The UV emission is coming from young, massive hot stars on the main sequence and hence is a direct probe of recent star formation in the tentacles of JO201. 
 While the stellar mass range probed by the UV is typically $M_{\star}>5 M_{\odot}$, with stellar lifetimes of 200 Myr or less, the $\mathrm{H}{\alpha}$ emission in star-forming regions is due to the recombination of hydrogen that is ionized by stars in the range $M_{\star}>20 M_{\odot}$, (O $\&$ B spectral type) whose lifetimes are 10-20 Myr or shorter \citep{Kennicutt_1998,Kennicutt_2012}. $\mathrm{H}{\alpha}$ and UV emission therefore probe different star formation timescales. The UV flux from the knots can thus have contributions from single or multiple bursts or prolonged star formation during the past few $10^8$ yr, while the $\mathrm{H}{\alpha}$ flux derives from the current "ongoing" star formation. Moreover, the $\mathrm{H}{\alpha}$ emission originates from the gas and therefore is an indirect tracer of ongoing star formation, while the UV probes directly the light coming from young stellar photospheres. 

In principle, interstellar shocks can also be responsible for generating UV radiation \citep{Shull_1979} and hence the UV emission from the stripped gas of JO201 can have a shock component. The MUSE data show that the radiation ionizing the gas throughout the disk and in the tails (except in the central region of the galaxy powered by the AGN \citep{Poggianti_2017a}) has emission-line ratios typical of ionizing radiation from young host stars (Bellhouse in prep.). Hence, it is reasonable to assume that the UV fluxes we observe with UVIT are dominated by young stellar light.
Note that the contribution of evolved population of metal poor extreme horizontal branch stars from the disk of the galaxy can contaminate the UV flux from the young stars, but this is expected to be negligible in the presence of strong ongoing star formation as in JO201. We stress that we are not using the UV flux from the central region of the galaxy in the present study as it is contaminated by the contribution from AGN. As we will show later, the star-forming origin of the UV emission in JO201 is also corroborated by the good agreement found using $\rm H\alpha$ and UV as independent star formation indicators.

\begin{figure*}\centering
\hbox{\includegraphics[width=0.50\textwidth]{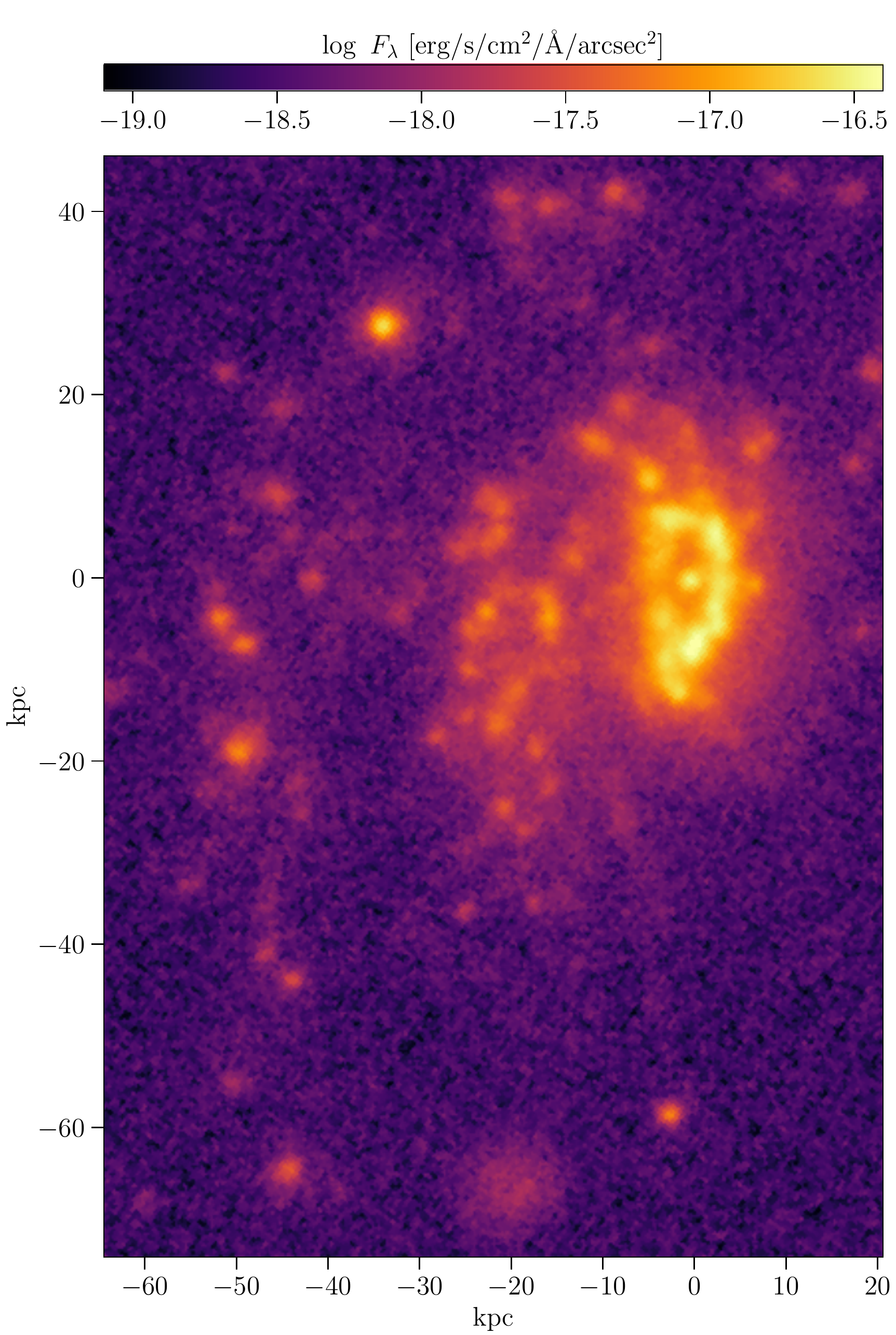}\includegraphics[width=0.50\textwidth]{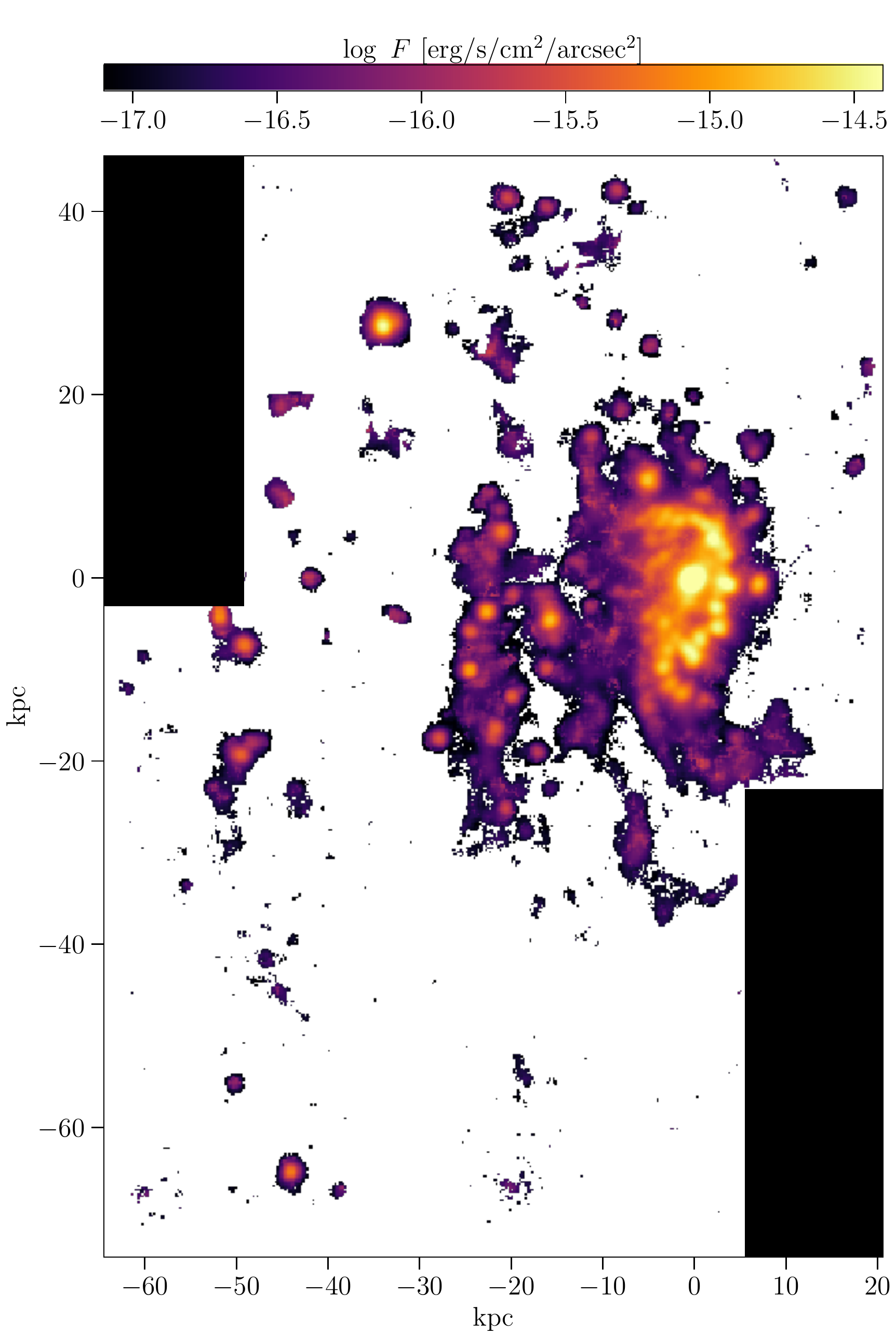}}
\caption{NUV (left) and $\mathrm{H}{\alpha}$ (right) image of JO201 with the same spatial scale and with a flux scaling that allows to bring out the the brightest pixels. The NUV image covers a larger field compared to the dust corrected $\mathrm{H}{\alpha}$ image.}
\label{figure:JO201nuvhalphaimage}
\end{figure*}

We compare the NUV and FUV images of JO201 with the $\mathrm{H}{\alpha}$ image in the following. The main motivation is to probe common features and also check for any missing features between the images. We compare the NUV and the $\mathrm{H}{\alpha}$ emission in Figure \ref{figure:JO201nuvhalphaimage}. We note that the MUSE plate scale is 0.2 $\arcsec$/pixel with resolution $\sim$ 0.7$\arcsec$  whereas the UVIT plate scale is 0.4 $\arcsec$/pixel with resolution of 1.2$\arcsec$ at the position of the galaxy. The image scaling is set such that the pixels corresponding to the brightest regions on both images are highlighted. Many features seen in the $\mathrm{H}{\alpha}$ image are also seen in NUV image.  There is generally a good coincidence between the NUV and $\mathrm{H}{\alpha}$ regions corresponding to peaks of star formation both in the stripped gas and on the disk of the galaxy. 

\begin{figure}
\includegraphics[width=0.50\textwidth]{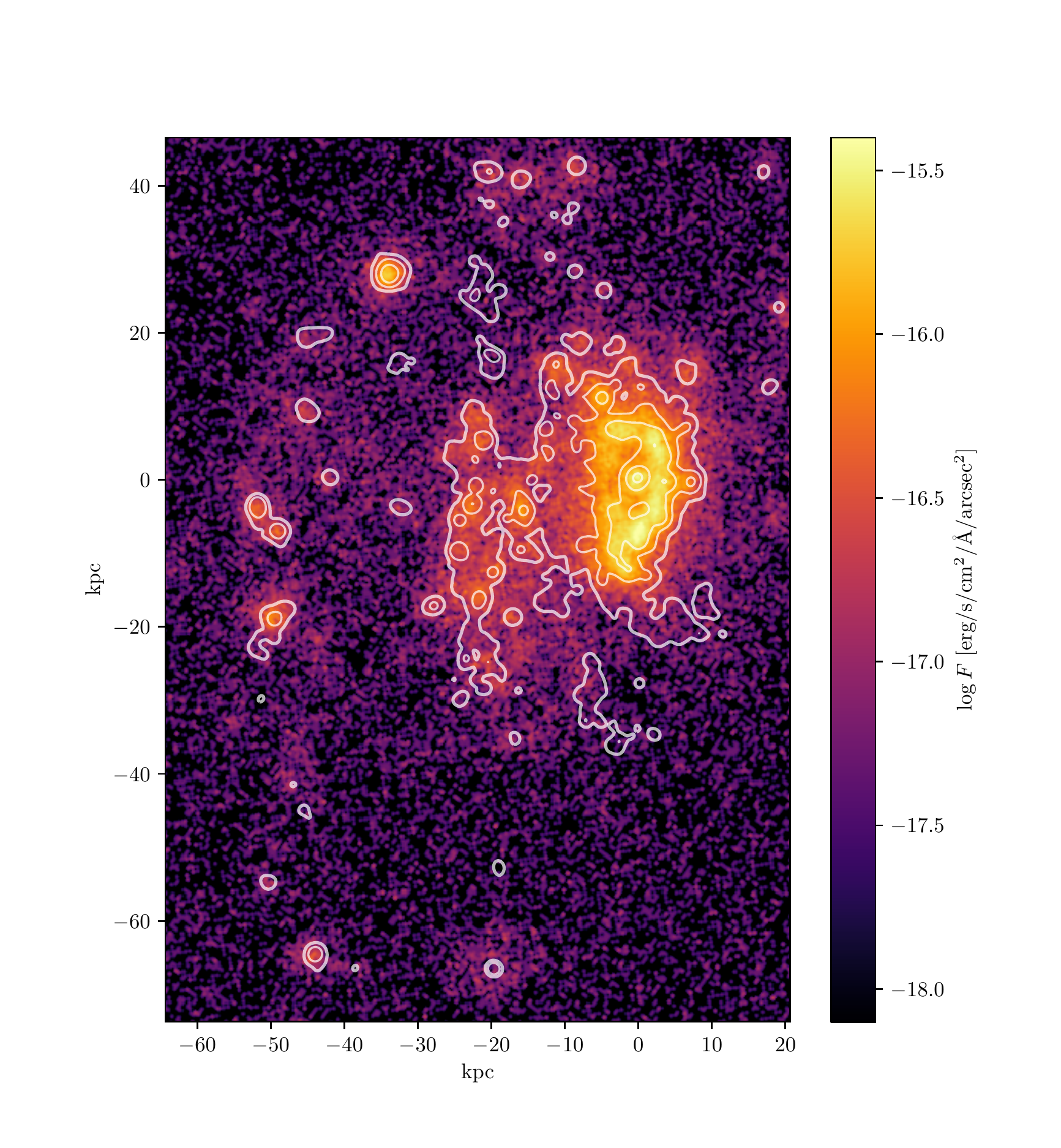}
\caption{The flux contours generated from the MUSE $\mathrm{H}{\alpha}$ image is overlaid over the FUV image of JO201.}\label{figure:JO201fuvimagehalphacontour}
\end{figure}

To better compare the two maps, in Fig \ref{figure:JO201fuvimagehalphacontour} we overlay the $\mathrm{H}{\alpha}$ contours over the FUV image. The overall coincidence between the two is remarkable. Nonetheless, there are a few regions relatively bright in $\mathrm{H}{\alpha}$ but not in NUV, and vice-versa. For example the regions in the north-east direction from the center of the galaxy (-20kpc and +30 kpc (X,Y)), the one between the galaxy and the outer 'arm' (-30 kpc and +15 kpc (X,Y)), and the one south of the disk (5 kpc and -20 Kpc (X,Y)) are certainly associated with JO201 (based on redshift information), but have no (or only some weak diffuse) NUV emission. Viceversa, the southern part of the eastern arm (-45 kpc and -45 kpc (X,Y)) shows a diffuse NUV region but a weak $\mathrm{H}{\alpha}$ counterpart. Interestingly, the regions with only $\mathrm{H}{\alpha}$ or only UV emission are generally those with the highest radial velocities in the $\mathrm{H}{\alpha}$ velocity map presented in \citet{Bellhouse_2017}, which are likely to be the furthest away along the line of sight, and those stripped first. Those with only $\mathrm{H}{\alpha}$ are very faint, and the lack of UV emission
can be due to the fact that the UV features may be below the detection limit of UVIT, though we point out that in principle such a discrepancy can also exists if there are only very young stars ($<$ 10Myr). Where only UV emission is detected, it is possible that star formation has already ceased therefore only stars with a low ionizing radiation are left. Finally, few of the other features seen in NUV are fore-ground or background objects ( -5 kpc and -60 kpc (X,Y)).

Note also that there is a general diffuse FUV emission in the intergalactic medium outside the $\mathrm{H}{\alpha}$ contour.  It is reasonable to conclude that the detected fluxes for overlapping regions from both images are having a common source of origin, which most likely is ongoing star formation.

\subsection{Star forming knots}

\begin{figure}
\includegraphics[width=0.3\textwidth]{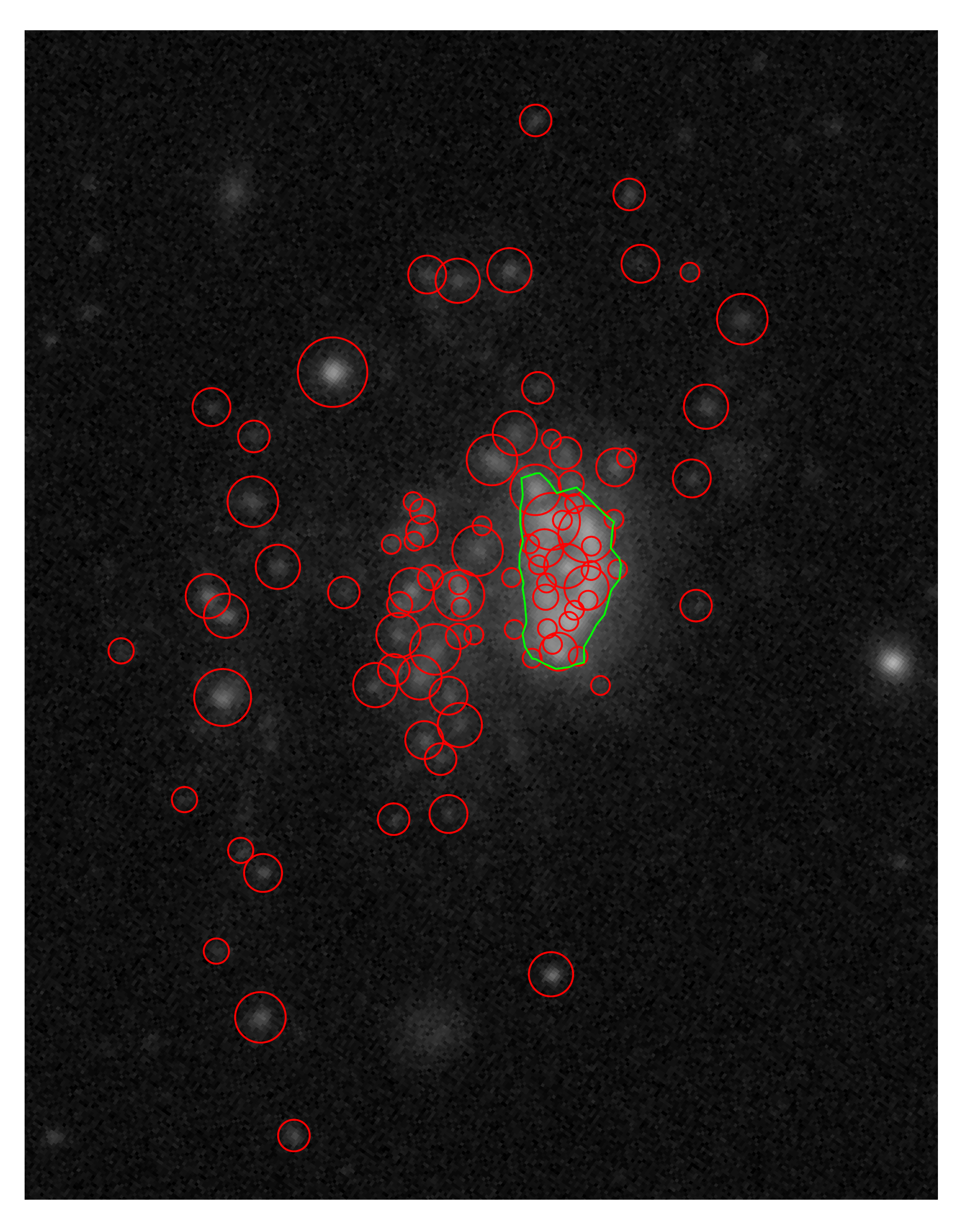}
\caption{The NUV grey scale image of JO201. The red circles indicate the 89 knots identified with the procedure described in the text. The stellar disk of the galaxy is shown with green contour and corresponds to the isophote (surface brightness $\sim$ 22.4 mag/arcsec$^2$) created from the optical $V$ band image.}\label{figure:JO201nuv}
\end{figure}

\begin{figure}
\includegraphics[width=0.55\textwidth]{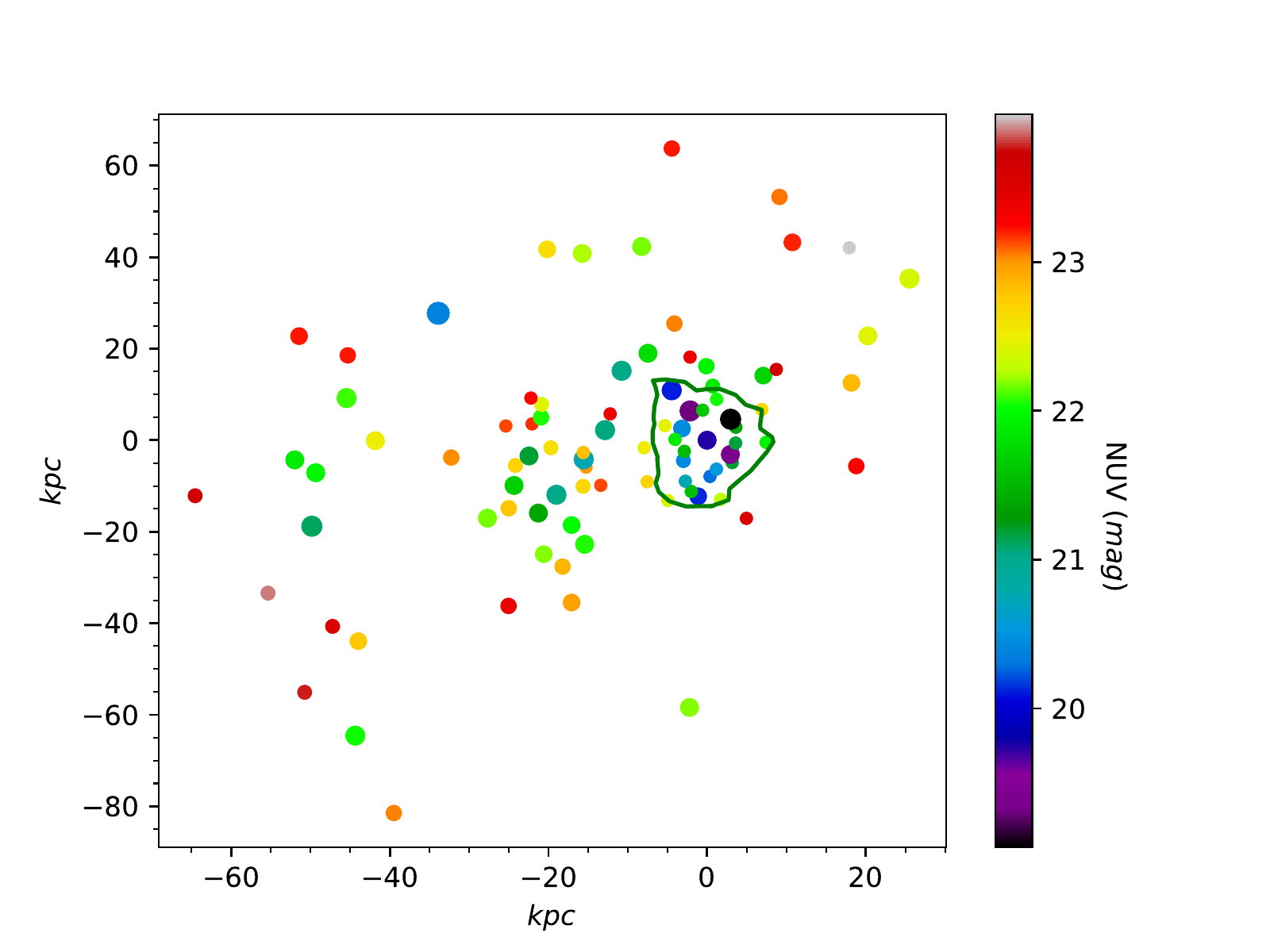}
\caption{The distribution on the sky of the detected knots with color-coded observed NUV magnitude. The scaling is shown in the color bar. The size of the point is proportional to the radius of the knot. Note the central bright knot, which is the emission from the active galactic nucleus. The region corresponding to the galaxy disk is shown with a green line.}\label{figure:Jo201nuvknotmag}
\end{figure}

The NUV knots are detected using a customized code in IRAF and FORTRAN, as described in \citet{Poggianti_2017}. The code was first run on the NUV image and we detected 89 knots of varying radii associated with the stellar disk and the tentacles on the NUV image of JO201. The MUSE data cover 85 of these knots. In principle, there can be background/foreground objects contaminating our bonafide knots belonging to JO201. In the following we will use the redshift provided by the MUSE spectra at each location to confirm that these knots indeed belong to JO201. Figure \ref{figure:JO201nuv} shows the detected knots (in red) overlaid over the NUV image of JO201. The stellar disk of the galaxy which corresponds to the isophote (surface brightness $\sim$ 22.4 mag/arcsec$^2$) created from the optical $V$ band image is shown with a green contour. We used the position and radius of these knots to measure the flux from both the NUV and the FUV images and the $\rm H\alpha$ image. NUV magnitudes are computed for the knots and are shown in Figure \ref{figure:Jo201nuvknotmag}. 

\begin{figure}
\includegraphics[width=0.5\textwidth]{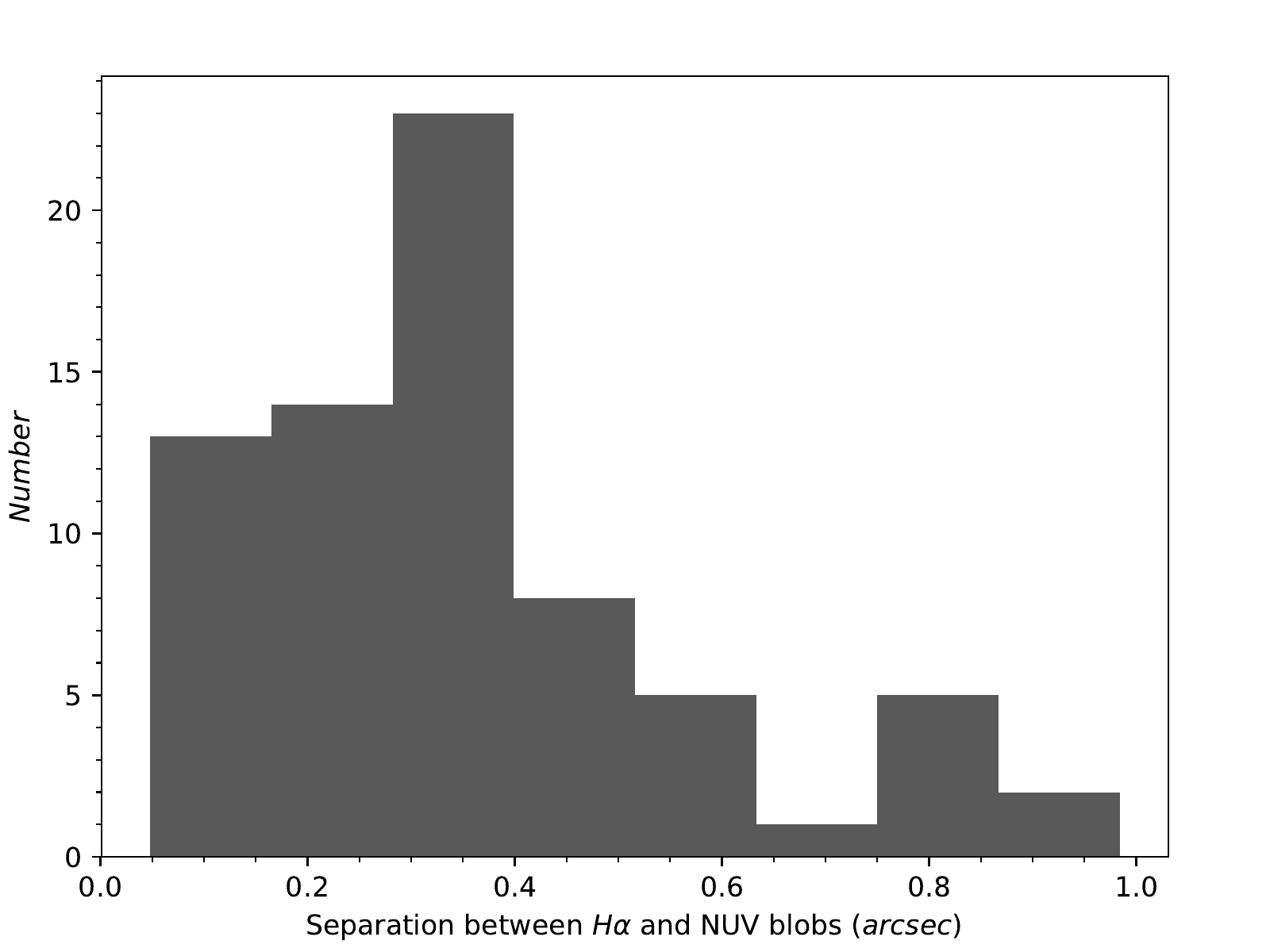}
\caption{Histogram of the offset (in arcsec) between the positions of knots detected from NUV and $\mathrm{H}{\alpha}$ images. No offset greater than 1 arcsec is present.}
\label{figure:NUV_halpha_knots_hist}
\end{figure}

The star formation in the knots outside the disk of JO201 can be interpreted as in-situ newly born stars from the ram-pressure stripped gas. Slightly older stars could be decoupled from the natal gas cloud and hence the UV (which directly traces < 200 Myr stars) and $\mathrm{H}{\alpha}$ (tracing < 10 Myr stars) emission could have an offset, as described for example in \citet{Kenney_2014} for the jellyfish galaxy IC3418 in the Virgo cluster. We checked for any offset between the UV and $\mathrm{H}{\alpha}$ peak emission within the knots of JO201. The knot detecting algorithm was independently run on NUV and $\mathrm{H}{\alpha}$  images. We present in Figure \ref{figure:NUV_halpha_knots_hist} the distribution of the relative offset in arcsec between the knots detected from NUV image and the knots detected from the $\mathrm{H}{\alpha}$ image. We found that the peak emission from the knots match within the instrument resolution of the images. No significant offset between the $\mathrm{H}{\alpha}$ and the NUV emission of the star forming knots in the tails is detected. This is somehow expected for since the galaxy is moving mostly in the direction of the observer and therefore any positional offset between the UV and $\mathrm{H}{\alpha}$ emission would be hard to detect in projection.

\begin{figure}
\includegraphics[width=0.5\textwidth]{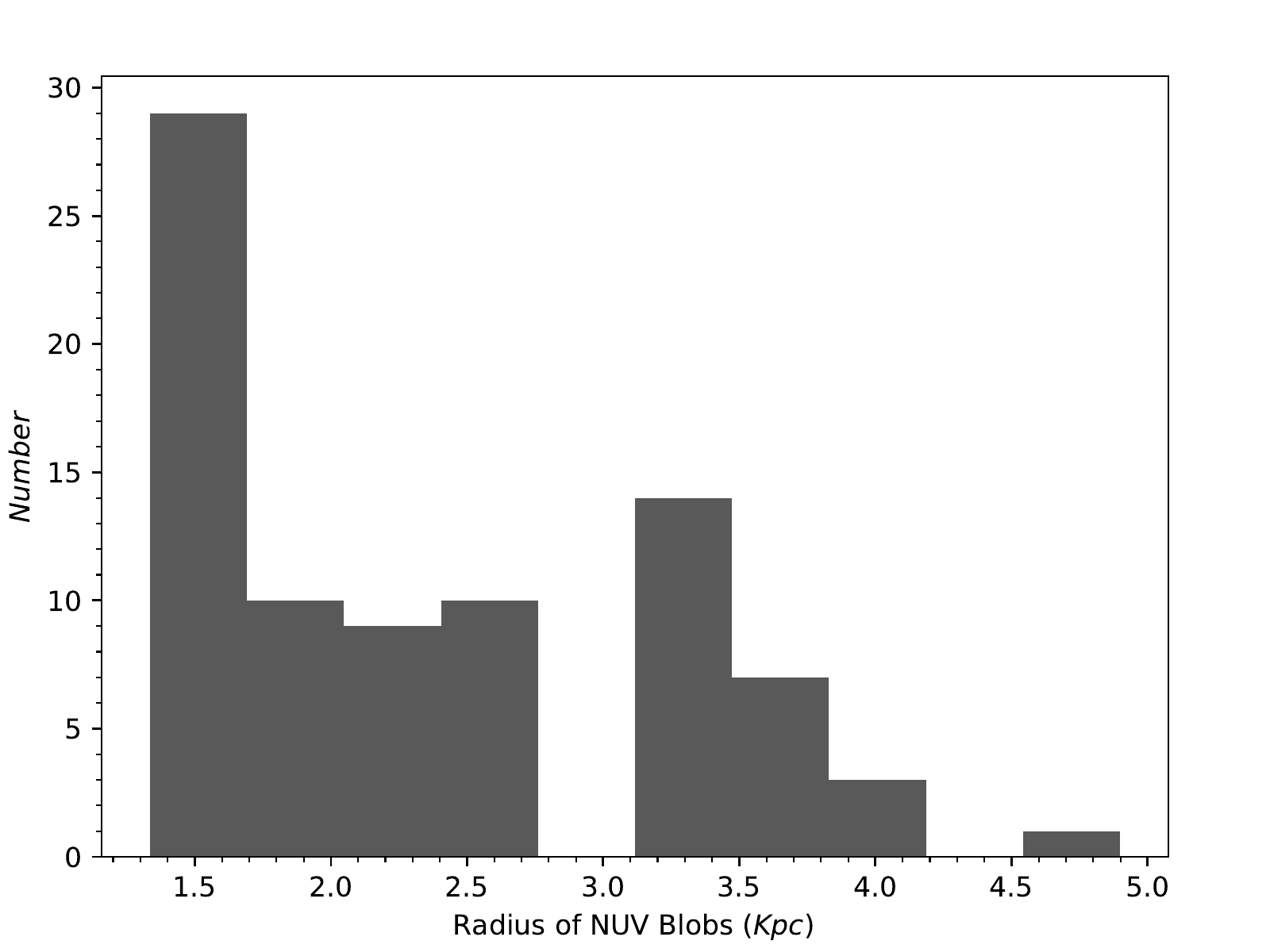}
\caption{The histogram of the radius of the star forming knots detected from the NUV image of JO201. The spatial resolution for the NUV image of JO201 is 1.3 kpc.}\label{figure:JO201knotradius}
\end{figure}

The size distribution of the knots detected from NUV imaging is shown in Figure \ref{figure:JO201knotradius}. The detected knots are of varying radius over 1.4 -to- 4.9 kpc. We note that the lower limit is set by the resolution of the NUV image, hence most of the UV knots might be in fact unresolved. Anyhow, for those that are resolved, we can conclude that the star-forming regions identified in the UV images can be as large as a $\sim$ 5 kpc in radius.

\subsection{Ultraviolet Extinction correction} 

\begin{figure}
\includegraphics[width=0.55\textwidth]{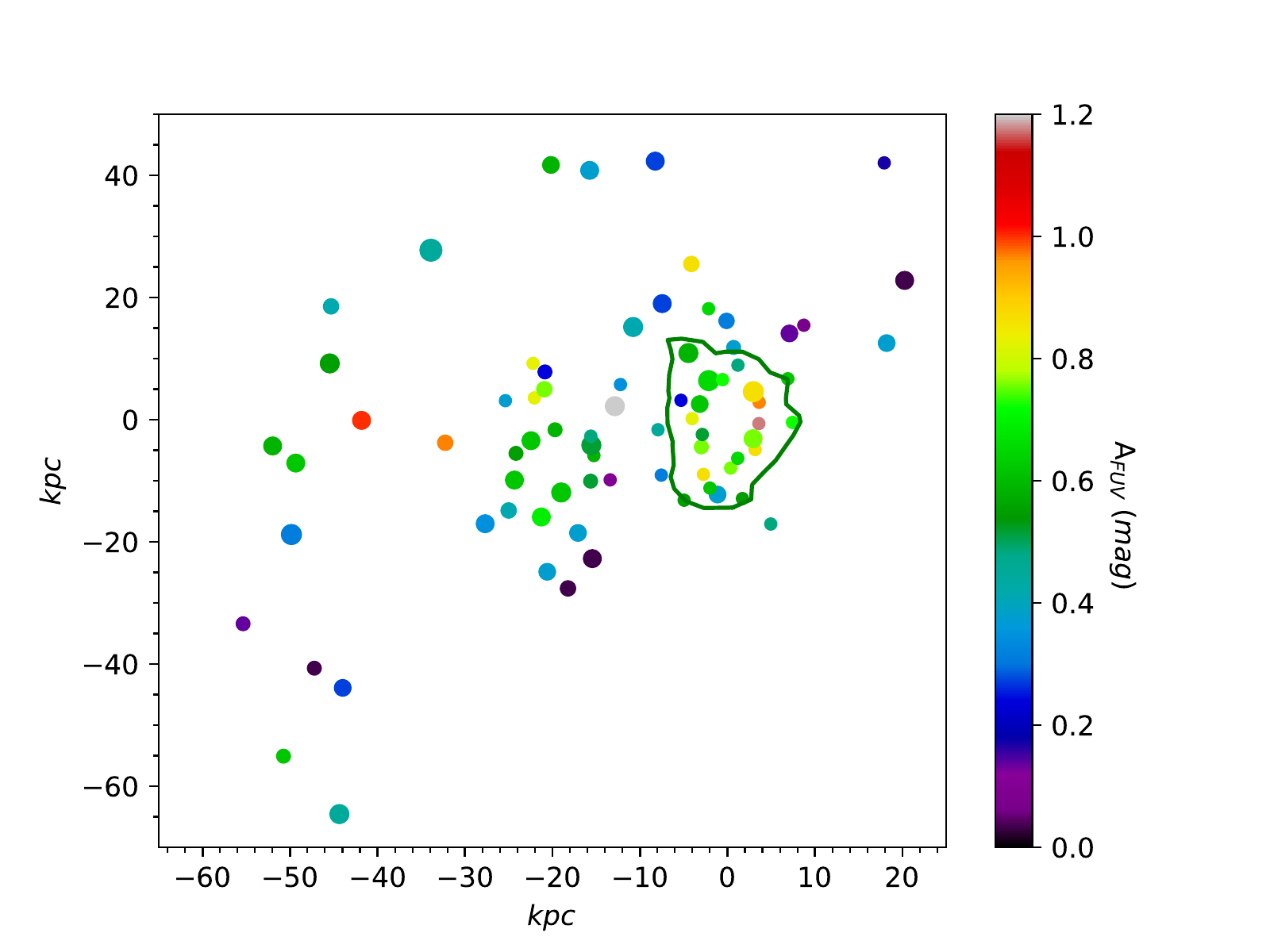}
\caption{The distribution of the FUV extinction values ($A_{FUV}$) of the knots as projected in the sky coordinate (J2000).The points are color coded for the $A_{FUV}$ as shown in the color bar and the size of the point is proportional to the radius of the knot. The region corresponding to the galaxy disk is shown with a green line.}
\label{figure:JO201nuvfuvext}
\end{figure}

To obtain the intrinsic UV fluxes of each knot from the observed fluxes we need to correct for extinction from dust both in our own Milky Way along the line of sight and within JO201. This is needed since star-forming regions are associated with significant amount of dust and the ultraviolet radiation is strongly affected by dust extinction. We use the FUV flux to estimate the star formation rate in the following section and hence correct the FUV flux for extinction. We first corrected the observed FUV flux for Galactic extinction (Av=0.0987 in the direction of JO201) applying the Cardelli extinction law \citep{Cardelli_1989}. We then corrected the observed FUV flux for rest frame extinction with the following method. We used the ratio of the $\mathrm{H}{\alpha}$ and $\mathrm{H}{\beta}$ emission line fluxes obtained from the MUSE data (Balmer decrement) assuming an intrinsic $\mathrm{H}{\alpha}$/$\mathrm{H}{\beta}$=2.86 and the Cardelli law to correct the $\rm H\alpha$ flux at each location (i.e.\ MUSE spaxel). From the comparison between the total $\rm H\alpha$ flux corrected and uncorrected for dust within each knot we computed the global $A_{\rm H\alpha}$ correction for each knot. According to the attenuation law from Calzetti et al. (2000), the ratio between the extinction at the FUV $A_{\lambda=1481}$ and $A_{\rm H\alpha}$ is equal to 1.38, therefore we used the appropriate value of $1.38 \times A_{\rm H\alpha}$ for each knot to correct the FUV flux. For those few knots which happen to fall outside of the MUSE field of view we could not apply any dust correction. Of the 85 UV knots which cover the footprint of the MUSE observations of JO201, 80 are confirmed based on redshift information as associated with JO201 and have a reliable estimate of $A_{\rm H\alpha}$. The FUV extinction values ($A_{FUV}$) of these 80 knots are shown in Figure \ref{figure:JO201nuvfuvext}. 

\begin{figure}
\includegraphics[width=0.55\textwidth]{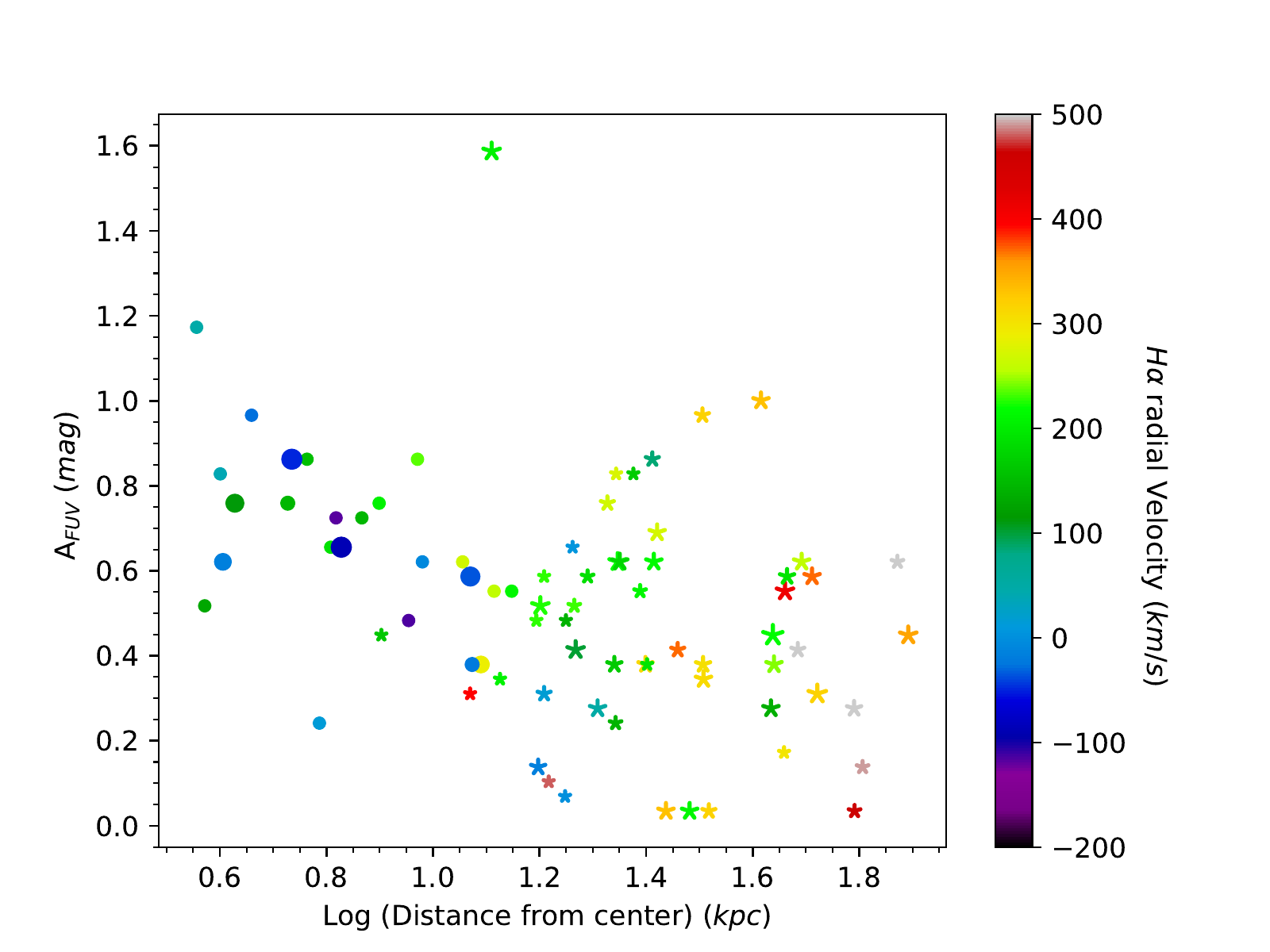}
\caption{The FUV extinction ($A_{FUV}$) value of the knots versus projected distance from the center of the galaxy. The points are color coded for the $\mathrm{H}{\alpha}$ radial velocity  as shown in the color bar and the size of the point is proportional to the radius of the knot. Circle markers correspond to the knots on the galaxy disk, while stars are in the tails.}
\label{figure:JO201nuvfuvextdist}
\end{figure}

We note that there might be a hint for the values of $A_{FUV}$ for the knots 
in the tails to be generally lower compared to those on the disk of the galaxy. This result is more evident in Figure \ref{figure:JO201nuvfuvextdist} which shows a decrease in the average $A_{FUV}$ value of the knots with projected distance from the center of the galaxy, although at large distances there is a very wide scatter. The knots are color coded according to the $\mathrm{H}{\alpha}$ radial velocity with respect to the center of the galaxy. The knots on the disk of the galaxy are shown in circles.

\subsection{Star formation rates}

\begin{figure}
\includegraphics[width=0.55\textwidth]{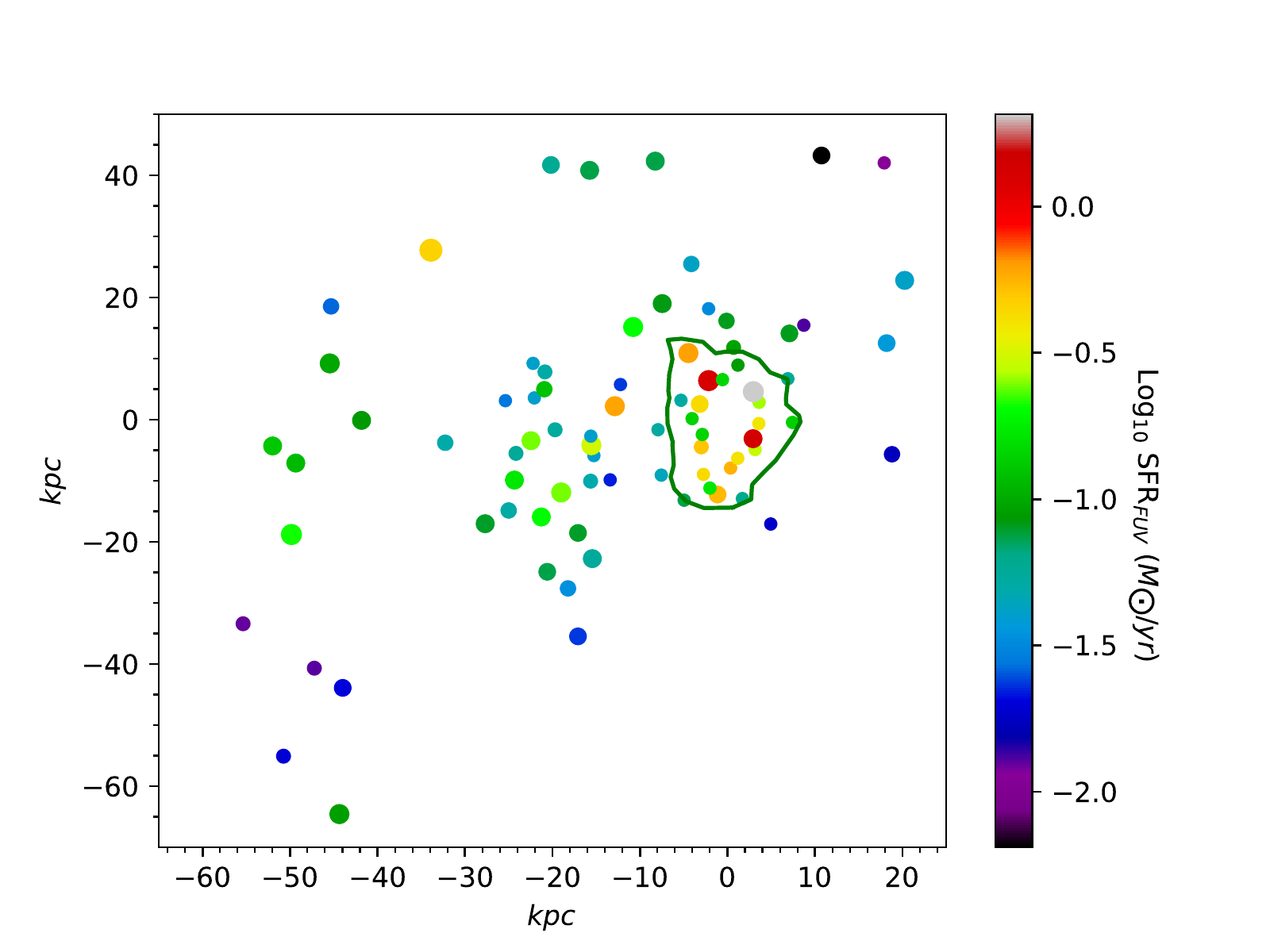}
\caption{The distribution of the star formation rates  of the star forming knots projected in the sky coordinate (J2000). The points are color coded for the log$_{10}$(SFR) as shown in the color bar and the size of the point is proportional to the radius of the knot. The region corresponding to the galaxy disk is shown with a green line.}\label{figure:JO201knotsfr}
\end{figure}

The dust-corrected FUV flux can be used to compute the star formation rate of star forming regions assuming a constant star formation rate over the past 10$^8$ years. The star formation rate along the tentacles and the disk of JO201 is computed for a Salpeter initial mass function from the FUV luminosity (L$_{FUV}$) \citep{Kennicutt_1998}. We used the following form of equation as described in \citet{Iglesias_2006} and adopted in \citet{Cortese_2008} to compute the star formation rate from FUV flux of the detected knots.  Note that the formula is derived using $Starburst99$ synthesis model \citep{Leitherer_1999} for solar metallicity and a Salpeter 0.1-100 $M_{\odot}$ initial mass function.

\begin{equation}
SFR_{FUV} [M_{\odot}/yr]  = \frac{L_{FUV}[erg/sec]} {3.83 \times 10^{33}} \times {10^{-9.51}} 
\end{equation}

\begin{figure}
\includegraphics[width=0.5\textwidth]{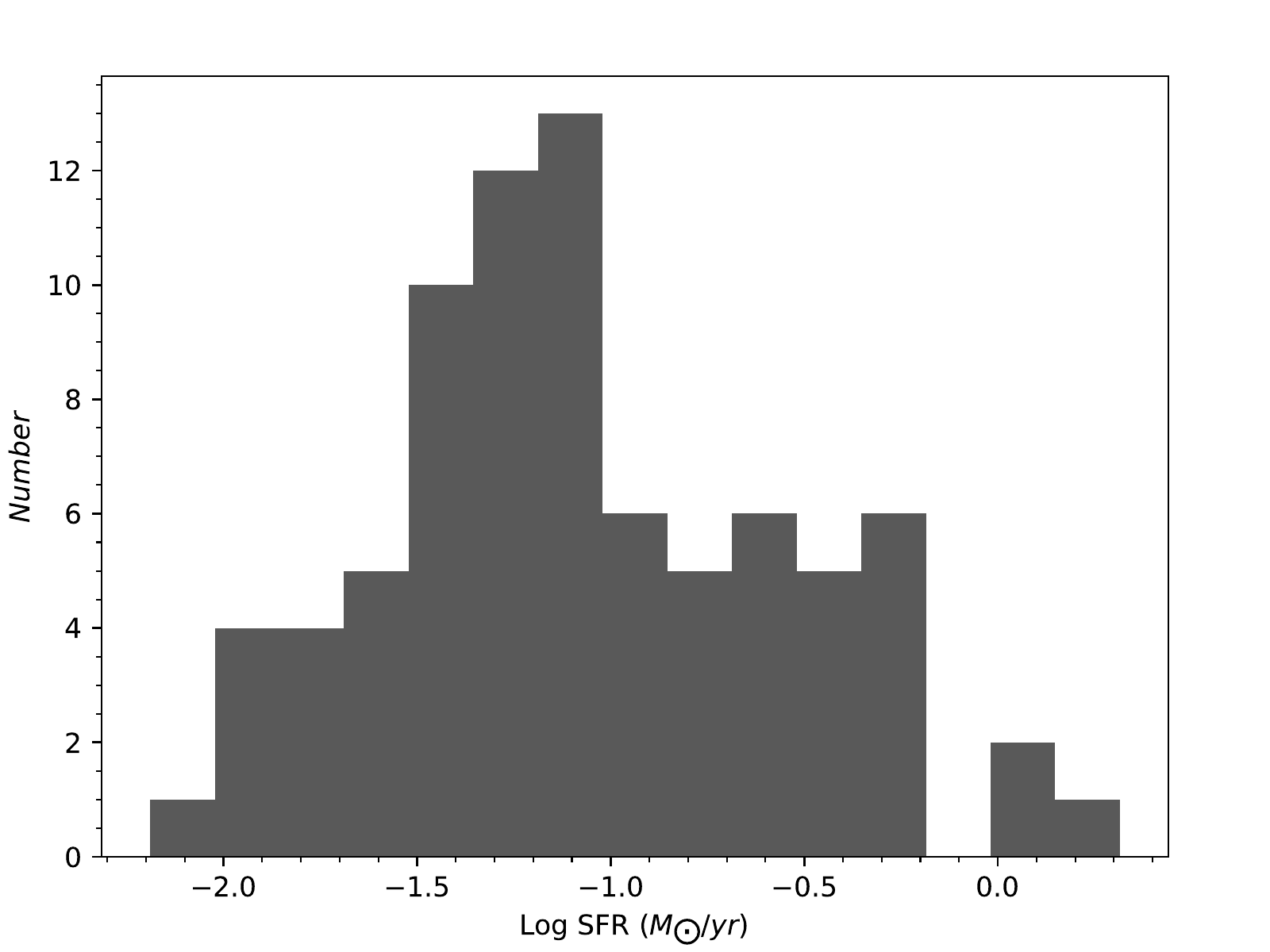}
\caption{The histogram of the star formation rates computed using the measured FUV flux of the star forming knots. Note that the star forming knots have a range in star formation rates.}\label{figure:JO201knotsfrhist}
\end{figure}

\begin{figure}
\includegraphics[width=0.5\textwidth]{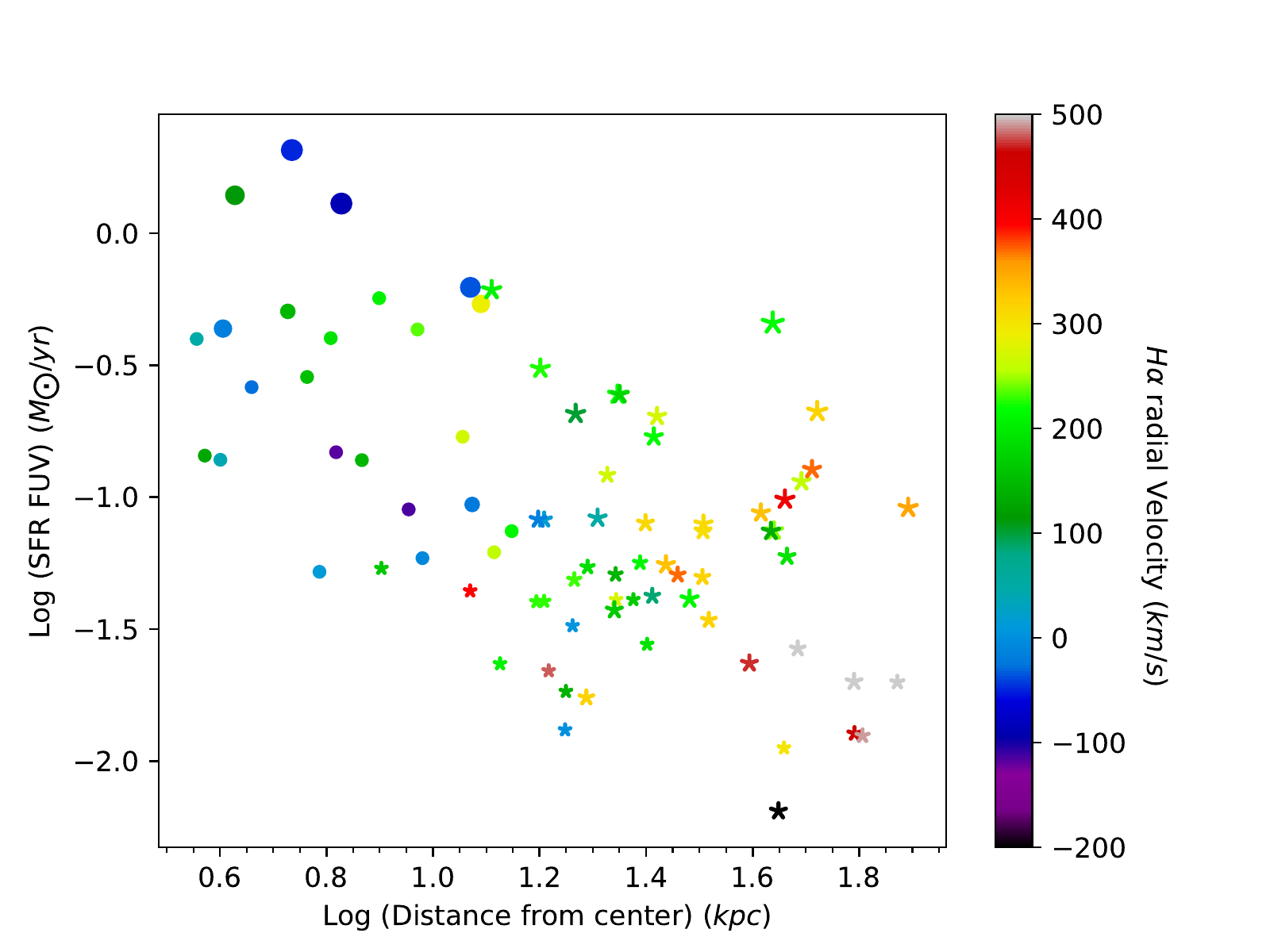}
\caption{The star formation rate of the knots decreases with the distance from the center of the galaxy.
The points are color coded for the $\mathrm{H}{\alpha}$ radial velocity  as shown in the color bar and the size of the points is proportional to the radius of the knot. Circle markers correspond to the knots on the galaxy disk, stars are knots outside of the disk. }\label{figure:JO201knotsfrdist}
\end{figure}

The sky projected position diagram of star formation rate of the knots is shown in  Figure \ref{figure:JO201knotsfr} and the distribution of the star formation rate of the knots is shown in Figure \ref{figure:JO201knotsfrhist}. The star formation rate of the knots decreases with distance from the center of the galaxy as shown in Figure \ref{figure:JO201knotsfrdist}.
The star formation rates of individual knots range from $\sim$ 0.01 to 2.07 $M_{\odot}  \, yr^{-1}$. 
We note that 24 knots are present on the disk and the remaining 56 are outside the disk within the intergalactic medium. The SFR of knots on the disk is having a median value of $\sim$ 0.3 $M_{\odot}  \, yr^{-1}$ and range from $\sim$ 0.05 to 2.07 $M_{\odot}  \, yr^{-1}$. The SFR of knots outside the disk is having a median value of $\sim$ 0.05 $M_{\odot}  \, yr^{-1}$ and range from $\sim$ 0.01 to 0.6 $M_{\odot}  \, yr^{-1}$. The highest star forming knot is located on the disk of the galaxy and the star formation rate is high for the knots on the disk and generally low outside the disk. The integrated star formation rate in the disk of the galaxy (summing up the individual SFR of knots) is found to be $\sim$ 10 M$_{\odot}$/yr and outside the disk is $\sim$ 5 M$_{\odot}$/yr. The total integrated star formation rate from FUV for JO201 is therefore $\sim$15 M$_{\odot}$/yr. For comparison, the values we find are much higher than the SFR values found in the tail of  the jellyfish dwarf galaxy IC3418 \citep{Hester_2010,Kenney_2014}. 

\begin{figure}
\includegraphics[width=0.5\textwidth]{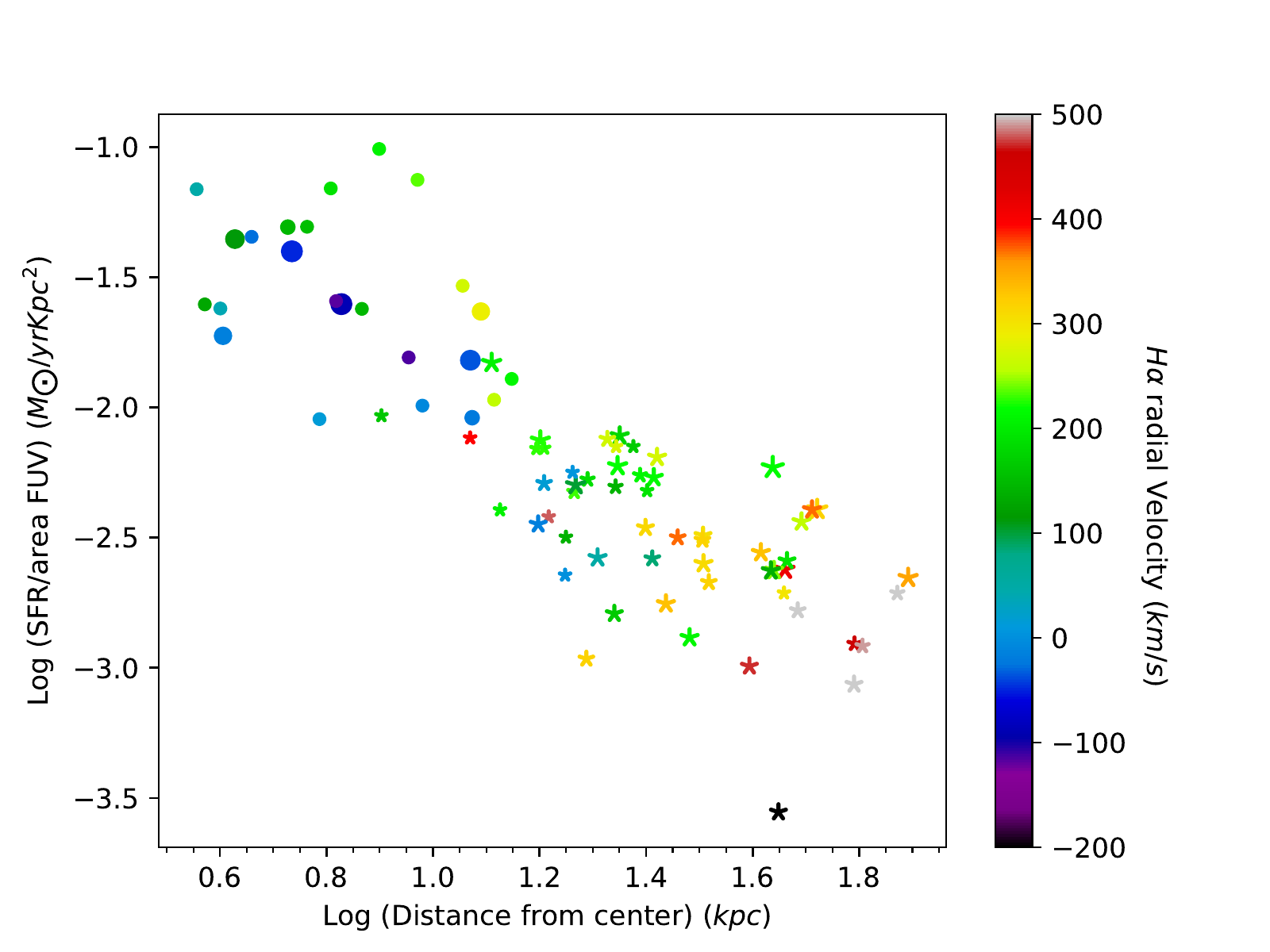}
\caption{The star formation rate density (SFR/Area) of the knots anti-correlates with the distance from the center of the galaxy. The points are color coded for the $\mathrm{H}{\alpha}$ radial velocity  as shown in the color bar and the size of the points is proportional to the radius of the knot. Circle markers correspond to the knots on the galaxy disk, stars are knots outside of the disk.}\label{figure:JO201knotsfrddist}
\end{figure}

We also computed the star formation rate density of the knots from the derived star formation rate and the area of the knots and searched for a dependence on the distance from the center of the galaxy. Figure \ref{figure:JO201knotsfrddist} demonstrates that the star formation rate density of the knots decreases with galactocentric distance.

We note that there are uncertainties in the projected distance due to the viewing geometry of the galaxy and the associated star forming knots in the sky plane. Figure \ref{figure:JO201knotsfrdist} and Figure \ref{figure:JO201knotsfrddist} paint a picture of the ram-pressure stripping process and the associated star formation in JO201. The knots forming stars far away from the disk of the galaxy are having lower star formation rates and star formation rate densities, and higher $\mathrm{H}{\alpha}$ radial velocity. The most distant knots must be hosting the gas that had been stripped first from the galaxy and thus moving with a higher velocity with respect to the disk of the galaxy. Also the gas phase metallicity of the star forming knots decrease with the distance from galaxy centre (Bellhouse et al in prep). This taken together implies that we are witnessing the ongoing star formation happening within the gas that has recently undergone an outside-in stripping event happening almost along the line of sight.

We also calculate the star formation rate of the NUV detected knots from the measured $\mathrm{H}{\alpha}$ flux with in the knots using the formalism described in \citet{Kennicutt_1998} and shown below.

\begin{equation}
SFR_{\mathrm{H}{\alpha}} [M_{\odot}/yr] = {7.9 \times 10^{-42} \times } {L_{\mathrm{H}{\alpha}}[erg/sec]} 
\end{equation}

The integrated star formation rate from the $\mathrm{H}{\alpha}$ emission within the NUV-detected knots is found to be $\sim$10 M$_{\odot}$/yr. The SFR of knots derived from the FUV and from $\mathrm{H}{\alpha}$ agree remarkably well (Figure \ref{figure:JO201SFRcomparison}). The values uncorrected for dust follow very closely the 1:1 relation (black points); when they are corrected for dust (red points), the UV-based SFR are slightly larger than those derived from $\mathrm{H}{\alpha}$. This was calculated assuming a stellar over gas E(B-V) value of 0.44 as in the standard Calzetti's formulation, while using a value of 0.32 the two estimates would follow the 1:1 relation. This might suggest that in the knots in jellyfish galaxies the difference between the extinction affecting the stellar light and the extinction of the youngest, most massive stars producing the ionizing radiation is even larger than in the average starburst local galaxies studied by \citet{Calzetti_2000}.

It is important to keep in mind that the star formation rates are obtained based on assumptions that are affected by large uncertainties, most importantly the shape of the IMF and the recent star formation history (bursty, continuous, or over what timescale). The star formation rate calculation is calibrated for galaxies which have ongoing star formation in the disk of the galaxy \citep{Kennicutt_1998}, while the jellyfish galaxy studied here is having star formation in a different environment. The estimates given above must therefore be taken with caution and considering all the possible caveats.

\section{Discussion} \label{sec:Discussion}

The star formation in the ram-pressure stripped tails of a few galaxies was studied at UV wavelengths using GALEX \citep{Chung_2009,Smith_2010,Hester_2010,Fumagalli_2011,Kenney_2014,Boselli_2018}. The  ultraviolet observation of JO201 in the Abell 85 galaxy cluster reveals a wealth of information on the ongoing star formation in a jellyfish galaxy. We detect star forming knots both on the galaxy disk and outside of the galaxy hanging in the intergalactic space. We are observing the triggered star formation in the gas that got stripped from the galaxy due to the impact of galaxy infall onto the hot intra cluster medium.  The UV observations presented in this paper provide strong evidence for ongoing star  formation in the  ram pressure stripped  gas of JO201. The knots of star formation detected from UV imaging are found to have significant $\mathrm{H}{\alpha}$ flux. This supports the notion that the UV and $\mathrm{H}{\alpha}$ flux of the knots are having a single origin and that is ongoing star formation. We note that CO APEX observations of JO201 \citep{Moretti_2018} have revealed the presence of significant molecular hydrogen (mass of H$_{2}$ $\sim$ 35.47 $\times$ 10$^{9}$ M$_{\odot}$) in the disk and the tails of the galaxy. The large reservoir of cold gas in the tail and the disk can be responsible for the vigorous star formation in the galaxy.

\begin{figure}
\includegraphics[width=0.55\textwidth]{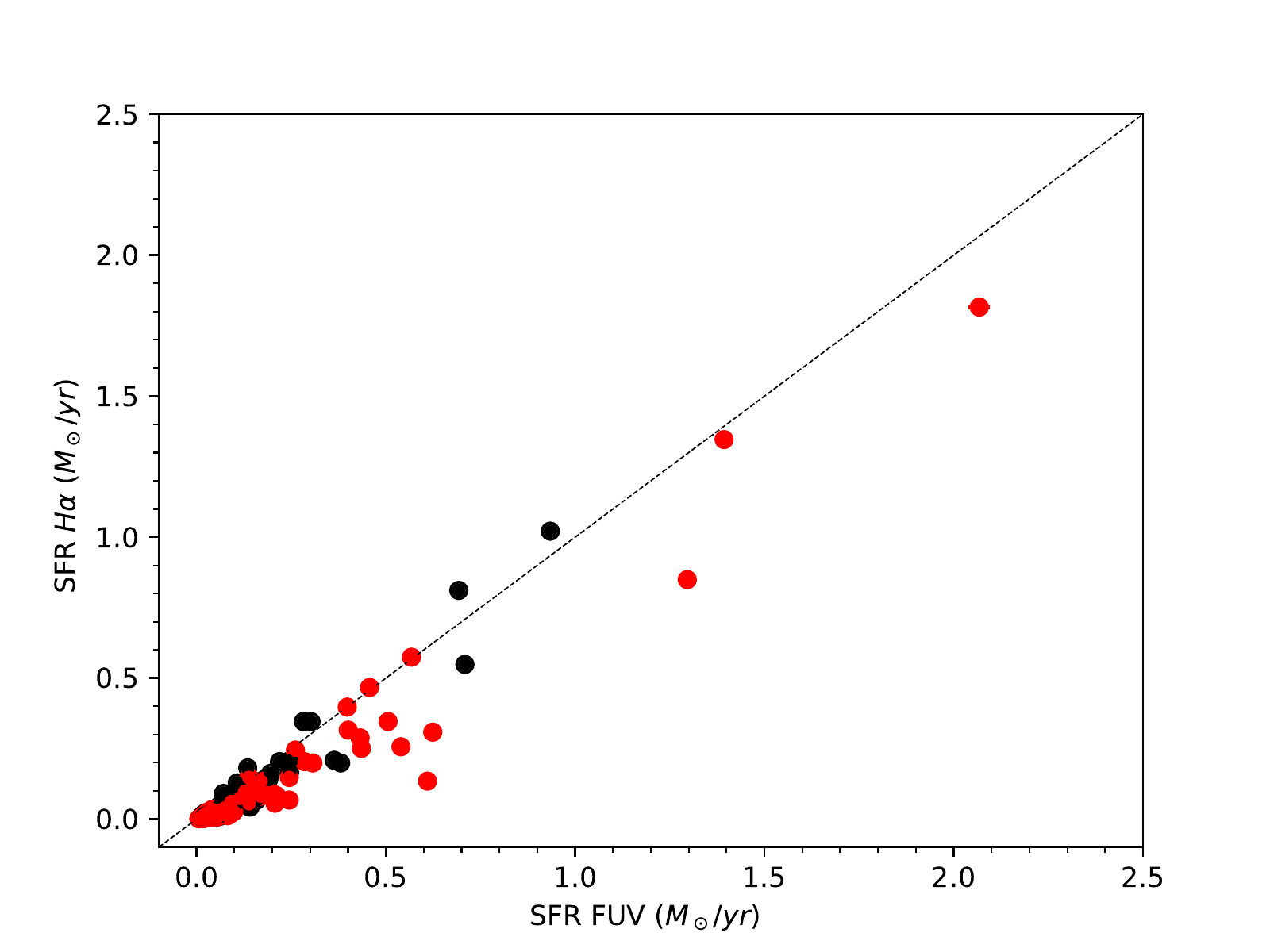}
\caption{The SFR derived from FUV flux is compared against the SFR derived from $\mathrm{H}{\alpha}$ flux for the 80 knots. The black points correspond to SFR derived with no extinction correction applied and the red points correspond to SFR with extinction correction applied to FUV and $\mathrm{H}{\alpha}$ flux values.}\label{figure:JO201SFRcomparison}
\end{figure}

We note that the peak emission of knots in UV and $\mathrm{H}{\alpha}$ images show very good correspondence in position. The offset seen in previous studies between the  $\mathrm{H}{\alpha}$ and UV emission of the knots is not seen in the present study. This can be due to the orientation at which the galaxy JO201 is falling into the cluster. As discussed in \citet{Bellhouse_2017}, JO201 is falling from the backside of the cluster with a slight inclination to the line-of-sight directed to the west. The projection of the knots along the line-of-sight may inhibit any offset between $\mathrm{H}{\alpha}$ and UV emission from the knots. Alternatively a lack of offset is intrinsic in nature to this jellyfish which can be dependent on the time scales involved in the stripping of gas from the galaxy and the onset of star burst in the knots. A statistical analysis of more jellyfish galaxies in UV and $\mathrm{H}{\alpha}$ is therefore needed to understand more details on the stellar emission regions. 

The star formation rate of the star forming knots is showing higher values on the galaxy disk (particular at the north western region) of JO201 compared to knots outside the disk in the intergalactic medium. The enhanced star formation rate on the disk can be interpreted as due to the trajectory of the galaxy falling into the Abell 85 galaxy cluster. The galaxy could be considered as freshly acquired from the field and might be undergoing the first infall into Abell 85 \citep{Bellhouse_2017, Jaffe_2018}. The enhanced star formation rate region (north western region) can be undergoing the first contact point of galaxy and the hot intra-cluster medium of Abell 85.

We found that the flux in the knots when not corrected for extinction at the source, yields a good agreement between the derived SFR$_{FUV}$ and SFR$_{\mathrm{H}{\alpha}}$ values as shown in Figure \ref{figure:JO201SFRcomparison}. This remarkably tight correlation deviates slightly more from the 1:1 relation (particular at lower values and for knots outside the disk of galaxy) when the correction for extinction is applied, possibly suggesting that the average dust extinction assumptions commonly used for local starburst galaxies do not apply to these systems. We note that in dwarf galaxies with SFR $\leqslant$ 0.1 M$_{\odot}$/yr in the local Universe, the SFR$_{\mathrm{H}{\alpha}}$ is found to be lower than SFR$_{FUV}$, similar to what is found here for the knots. In dwarf galaxies this effect is attributed to variations in the stellar IMF creating a deficiency of high mass stars \citep{Lee_2009}, and this is a possibility also in our case.

The integrated SFR for JO201 derived from the FUV flux (SFR$_{FUV}$ $\sim$ 15 M$_{\odot}$/yr) is found to be rather typical for normal star forming galaxies of this mass in the local Universe. The specific star formation rate (SSFR) of JO201 is calculated to be $10^{-9.6} yr^{-1}$ and falls on the star formation main sequence in the SSFR $vs$ stellar mass plot of star forming galaxies in the local Universe \citep{Salim_2007,Bothwell_2009,Paccagnella_2016}.

Finally we note that the UV study of JO201 presented here can be considered as a bench mark for observing higher redshift jellyfish galaxies using big optical telescopes. The rest frame UV emission will then be redshifted to  optical wavelengths. The clumpy star formation in the intergalactic medium can also be speculated to be similar in nature during the peak of star formation epochs at $z > 2$. The study of such systems in UV in the local Universe can give more insights into the triggered star formation in dense environments.

\section{Summary}\label{sec:Summary}

We have studied star formation in the jellyfish galaxy JO201 (taken from GASP sample) using the ultraviolet imaging observation from UVIT. Intense star formation is seen in the tentacles and disk of the galaxy, in agreement with the molecular gas detections in the JO201 we present elsewhere \citep{Moretti_2018}. We compared the FUV/NUV imaging data with the  $\mathrm{H}{\alpha}$  imaging data of JO201 and following inferences are made.

\begin{itemize}
\item The tentacles and main body of the galaxy show strong UV emission. The emitting regions in UV and  $\mathrm{H}{\alpha}$ are showing remarkable correlation. $\mathrm{H}{\alpha}$ emission is originating from the hot ionized HII regions surrounding the young OB stars, where as UV is coming directly from the photospheres of OBA stars. We confirm that both emissions have the same origin and that is ongoing star formation.

\item We search for a possible (physically motivated) spatial offset between the UV and $\mathrm{H}{\alpha}$ emission from the detected knots, but could not detect any offset above the instrument resolution.

\item We detected and confirmed 80 star forming knots on the disk and tentacles of galaxy JO201 from the NUV imaging data. The FUV extinction for the knots  are computed making use of the A$_{V}$ values derived from the Balmer decrement.

\item The star formation rates of individual knots are derived from the extinction corrected FUV flux and found to range from $\sim$ 0.01 to 2.07 $M_{\odot}  \, yr^{-1}$. We show that the star formation rate of the knots derived from FUV flux agree very well with the ones derived from $\mathrm{H}{\alpha}$ flux.

\item Both the star formation rate and the star formation rate density of individual knots decrease with the distance from the center of the galaxy.

\item The integrated star formation rate for JO201 derived from FUV is $\sim$ 15 M$_{\odot}$/yr and is shown to be comparable to star forming galaxies of similar mass range in the local Universe.

\item We demonstrate that our unprecedented deep UV imaging study of the jellyfish galaxy JO201 show clear signs of extraplanar star-formation activity, resulting from a recent/ongoing gas stripping event.\\

The study of JO201 like systems in UV in the local Universe can give more insights into the triggered star formation in dense environments at high redshifts.

\end{itemize}

\section*{Acknowledgements}
We would like to thank the anonymous referee for constructive comments that have helped improving the scientific content of the paper. This publication uses the data from the AstroSat mission of the Indian Space Research  Organisation  (ISRO),  archived  at  the  Indian  Space  Science  Data Centre (ISSDC).  UVIT  project  is  a  result  of collaboration  between  IIA,  Bengaluru,  IUCAA,  Pune, TIFR, Mumbai, several centres of ISRO, and CSA. Based on observations collected by the European Organisation for Astronomical Research in the Southern Hemisphere under ESO program 196.B-0578 (MUSE). We acknowledge financial support from PRIN-SKA 2017. Y.J. acknowledges support from CONICYT PAI (Concurso Nacional de Inserci\'on en la Academia 2017), No. 79170132. B.V. acknowledges the support from an Australian Research Council Discovery Early Career Researcher Award (PD0028506).This research made use of Astropy, a community-developed core Python package for Astronomy \citep{Astropy_Collaboration_2018}.












\bsp	
\label{lastpage}
\end{document}